body of paper

\documentstyle[preprint,prl,aps,epsfig]{revtex}
\begin{document}
\preprint{}
\draft
%
%
\title{Stroboscopic Laser Diagnostics for Detection of \\
Ordering in One-Dimensional Ion beam.}
\author{R. Calabrese{\dag}, V. Guidi{\dag}, P.Lenisa{\dag},
 E. Mariotti{\S}, L. Moi{\S} and
 U.Tambini{\dag}}
\address{\dag Dipartimento di Fisica dell'Universit\`a, I-44100 Ferrara, Italy;
        INFN - Sezione di Ferrara, Italy.\\
\S Dipartimento di Fisica dell'Universit\`a, I-53100 Siena, Italy.}
\date{\today}
\maketitle
%
%
\begin{abstract}
A novel diagnostic method for detecting ordering in
one-dimensional ion beams is presented. The
ions are excited by a pulsed laser at two different positions along
the beam and fluorescence is observed by a group of four photomultipliers.
Correlation in fluorescence signals is
firm indication that the ion
beam has an ordered structure.
\end{abstract}
%
%
%
%
\narrowtext
\section{Introduction}
In the past few years the interest in the field of ion crystallization
has become increasingly stronger from both theoretical and
experimental points of view. This research effort has led to the
successful detection of crystallization in an ion trap \cite{[1]} and in a
mini-quadrupole storage ring \cite{[2]}. In these systems, ions are at rest
in the laboratory frame and the transition to the ordered state is
achieved by laser cooling \cite{[3]}.

	At the same time, the possibility has been suggested to cool an
ion beam in a storage ring to the extent that a crystalline phase is reached
even in
such system, in which particles travel at high velocity \cite{[4]}. In this
case, compaction of phase-space of the ion beam is obtained by
electron \cite{[5]} and laser cooling techniques. Recently, an ion storage
ring dedicated to crystallization studies has been proposed \cite{[6]}.

	Among the items one should consider in designing such a ring is
the problem of cooling the ion beam and the diagnostics of the
ordered state. It is known that an ion beam can be efficiently laser-cooled
only longitudinally but there is no experimentally tested technique that
allows cooling of the transverse degrees of freedom of the ion beam as
efficiently as longitudinally. Although some proposals for an efficient
transverse cooling are under consideration, the lack of such a technique
is a severe obstacle to the reaching of any ordered state.

	The diagnostics of the ordered state is another crucial point, because
a clear and unambiguous detection of the crystalline state is
necessary to validate experimental findings. For the case of ions in
traps, one can use a CCD-based detection since ions are practically at
rest \cite{[2]}.

	Unfortunately, to our knowledge, no satisfactory method to
detect ordering of an ion beam by resolving the single ion
 has been proposed so far. It
has been suggested that Schottky-noise-based pick-ups might
detect the passage of the single particle of the beam but, in a
storage ring, the bandwidth of such an instrument would be greater
than $100$ GHz, that is by far higher than any available device can achieve.
 Among
optical methods, some possibilities are described in Ref.\cite{[7]}; these
instruments should provide a response correlated with ion order
but they would never allow a direct and unambiguous evidence of the
position of the single ion.

	In this paper we propose a fluorescence-based method to
detect ordering in a one-dimensional ion beam.
\section{Basic principles of the method}
A pulsed laser - resonant with
the travelling ions - is split in two parts, which simultaneously
cross the ion beam at right angle at two nearby positions along
the storage ring (see Fig.1). This laser-to-ion crossing area is
followed by four photomultipliers which detect the photons emitted by
the ions that have previously been excited by the laser beams.
 The signals recorded by the photomultipliers
are analyzed when
one laser beam is moved with respect to the other one. In the absence of
ordering,
 no correlation in fluorescence signals should be
recorded while changing the relative distance of the two laser beams.
 On the contrary, if a string were obtained as a result
of cooling, a strong correlation between the signals should be
observed.
Suppose that one of the four photomultipliers detects the fluorescence
of an ion excited by one of the laser beams. If one of the other three
photomultipliers detects a simultaneous fluorescence signal, it means that
the other laser has interacted with another ion in the string, in turn
indicating that
the distance between the two ion-to-laser crossing points is an
integer multiple of the string interparticle spacing. Then, by slightly
moving the second laser beam, the correlation signal should vanish.
A sort of periodical dependence on the distance between the laser beams
should appear, whose contrast will mainly depend on the
detection efficiency.

	The diagnostics is conceived with the aim of
detecting ordering within a string of ions in a storage ring. When an
ion beam in a storage ring undergoes sufficiently strong cooling, a
phase transition to an ordered state is expected to occur \cite{[Schiffer]}.
The simplest ordered structure is a one-dimensional string.
For this system the degree of ordering increases
as temperature decreases without any sharp transition as is the case for
three-dimensional systems \cite{[Land]} \cite{[Emer]}.
Typical values of the interparticle spacing for the string configuration lie
between $s=10-100~\mu$m; for the following we shall assume $s=50~\mu$m.
At non-zero temperature, ions are expected to oscillate both in the transverse
and longitudinal directions (incoherent motion). Moreover, for
a string longitudinal oscillations of equilibrium positions can also
occur (coherent motion) through long-range waves.
For the diagnostic system concerned through this paper the distance between
the two laser beams can be chosen to match interparticle spacing of the string.
In this case, the effect of coherent motion is uneffective as the
nearest-neighbor spacing is relatively uniform.
The effect of long wavelength on interparticle oscillations is negligible for
a relative distance of a few lattice steps. An analytical evaluation of this
 effect can be
found in Ref. \cite{[Avilov]}. Short wavelengths mostly affect the fluctuations
in interparticle spacing, $\delta$$s$. Based on Ref. \cite{[Avilov]}, a rough
 estimate holds:
 $<\delta$$s^{2}>$/$s^{2}$$ \simeq$  1/$\Gamma$,
 where $\Gamma$ is the plasma parameter for the ion beam.
When the ion beam is being cooled $\delta$$s$ becomes even lower than $s$;
as an example, at
$\Gamma=100$, $\delta$$s$ = $5\mu$m.
If the distance between the laser beams were larger than one interparticle
spacing, the correlation signal should become progressively weaker.

	The configuration of the diagnostic device can also compensate
for a possible influence of transverse oscillations. Since the laser beams
cross the string orthogonally and their spots are small enough that they
 do not overlap, the ions suffering
transverse oscillations in the direction of the laser can always be resonant,
irrespective of their coordinates along that axis. Ion oscillations in
direction orthogonal to both the laser and the string could in principle
 move the target ion outside the laser beam spot. To avoid this effect - and
the consequent less efficiency - the laser beam can be focused by a cylindrical
lens. This optical element can be arranged to produce a focal segment
orthogonal to the directions of both the laser and the string. In this way
 the locations where the laser beams impinge on the string are two thin
regions; these can be as wide as several hundreds of microns in one dimension
 without overlap between them. Transverse oscillations of
a very cold ion beam are expected to be lower than this value.

\section{An application of the method to a real storage ring}
	We shall discuss a possible implementation of such diagnostic device
 with reference to the case of a $ ^{24}\!Mg^{+}$
string. In order to provide an example of a possible application, we
shall refer to the ASTRID Storage Ring at Aarhus: its main
parameters can be found in Ref. \cite{[Nielsen]}. With the above
assumptions, the velocity of the ion beam is about
$ v=8.97\cdot10^{5}$ m/s.
The time duration of each laser pulse must be much shorter than
the time, $T=s/v=56$ ps, taken by an ion to travel one interparticle
spacing. On the other hand, the laser pulse must not be too short because this
would lead to a very broad frequency pattern, in turn making more difficult
the filtering between laser photons and fluorescence photons, as discussed
below. A laser whose pulses are 2 ps long meets these requirements.

 In the following we shall refer
to a commercially available, frequency tripled  $ Ti:Al_2O_3 $ pulsed laser
(50 MHz repetition rate)
, with some nJ/pulse.
Considering the finite duration of a Gaussian-shaped laser pulse, the
probability to excite a $ ^{24}\!Mg^{+}$ ion is about 0.5 for
the transition under consideration ($ 3 s ^2\!S_{1/2} \to 3 p ^2\!P_{1
/2}$).
 The laser frequency must
be resonant with the ion's transition energy in its reference system
($\lambda$=279.6 nm).
 The
laser beams need focusing to spot much smaller than the interparticle spacing
($50~\mu$m). This can be done since it is experimentally possible to focus a
laser beam within $5~\mu$m (FWHM). The laser focusing systems need to be placed
in
the vacuum chamber and should be movable in order to avoid interference with
the ion beam during normal operation of the storage ring.

The deacay region is viewed by four photomultipliers, located each behind
its own window. The window length ($2L$) matches the decay length for
the ion de-excitation. The four windows cover about $50$$\%$ of the azimuthal
 acceptance of the fluorescence photons. The distance $d$,
 between the upstream edge of photomultiplier acceptance and
 the interaction region between the ion and the
laser beam should be as short as possible (see Fig. 2). We assume $d=10$ mm,
$2L=20$ mm, a beam-pipe radius, $R=35$ mm, and the lifetime of
the upper level, $\tau=3.5$ ns.
  The photomultipliers are single-photon
detectors, which can be assumed to have a quantum efficiency
of $23$$\%$ and a rate of background counts of about $100$ c/s.

The signals from the four photomultipliers are discriminated ($20$ ns signal
 width) and the logical signals are ANDed two by two to form 6
combinations. These are then ORed; a positive logical level for the OR is an
event in which the two ions, excited by the laser beams,
 have both emitted a photon.

Each
photomultiplier must be equipped with a filter to intercept stray laser
 photons. Let $\theta$ be the angle
between the direction of the ion beam and the direction of a photon
emitted by an ion.
Due to the finite duration of a laser pulse (2 ps), the frequency spectrum
is of the order of
500 GHz. Considering that the laser beams impinge on the string at right angle,
 the photons of the lasers are overlapped in frequency with those from
natural fluorescence between the angles $\theta=85^{0}$ and $\theta=95^{0}$.
Filters are designed in order to discard all photons impinging with an
 angle $\theta
 > 70^{0}$. The geometrical acceptance of the system allows detections of
photons only for $\theta > 50^{0}$. This corresponds to a lower limit for
the filter bandwidth of 0.25 nm.
The distance between the two laser beam spots can be varied by moving the
 optical set-up; this movement can be done with an accuracy better
than the range of longitudinal oscillation for interparticle spacing.

	As described above, by moving the second laser beam with respect to
the first one
and counting the coincidence events of the photomultipliers,
 one can assess if the distance between the two laser-to-ion crossing points
is an integer multiple of the string's interparticle spacing.
After having met this condition, by slightly moving
the second laser, one would record only accidental coincidences;
 this is true since, if the first laser
excites an ion, then the second laser beam can no longer be synchronous
with any other ion in the string. Appearance of correlation in the
signals with the same periodicity of the string would be firm
indication of ordering in the beam.

	The total probability that one of the four detectors sees
the fluorescence of an ion excited by one of the lasers is
about $9.3 \cdot 10^{-4}$. Considering a purely random jitter of the laser, a
laser beam spot of $5~\mu$m and a typical interparticle spacing of $50~\mu$m,
the probability that a laser pulse crosses an ion in the beam can be
roughly estimated as $1/10$.

	Therefore, the probability that a simultaneous de-excitation
will be recorded by the device is $1 \cdot 10^{-6}$. Considering a repetition
rate of the laser of $5\cdot10^7$ Hz, one expects a counting rate of
approximately 50 Hz. When the distance of the laser-to-ion beam
crossing points is not an integer multiple of the string's step, the rate
of accidental coincidences is estimated to be $0.25$ Hz. On the contrary,
a non ordered ion beam would exhibit a coincidence rate of $10$ Hz
independent of the relative positions of the laser beams.

	In order to check these results, we have developed a Monte Carlo
simulation. Ion oscillations around
their equilibrium positions, the excitation process of an
ion by the laser light, the spontaneous emission, the geometrical
acceptance and efficiency of the detector (filters +
photomultipliers) are taken into account. Fig.3 shows the counting rate
 of coincidences
versus the position of the second laser. A strong
correlation signal is achieved for the case of an ordered string.

\section{Conclusions}
A novel method to detect ordered structures of an
ion beam in an unambiguous way is proposed, and its feasibility is
demonstrated. The method provides firm observation of ordering within the ion
beam using available technology.
This could be
profitably applied to operating storage rings.
\acknowledgements
The authors are grateful to A. Burov, R. W. Hasse, S. Gustafsson,
L. Piemontese, and L. Tecchio for
 critical reading of the manuscript.
%
%

%
%
\begin{figure}
\caption{A sketch of the experimental apparatus: a) general view, b)
side-view.}
\end{figure}

\begin{figure}
\caption{A sketch of the ion-laser interaction region.}
\end{figure}

\begin{figure}
\caption{Counting-rate of coincidences versus second laser position
 for a string and for a hot beam with the same density .
The acquisition time of events for the Monte Carlo is $1$ s.}
\end{figure}
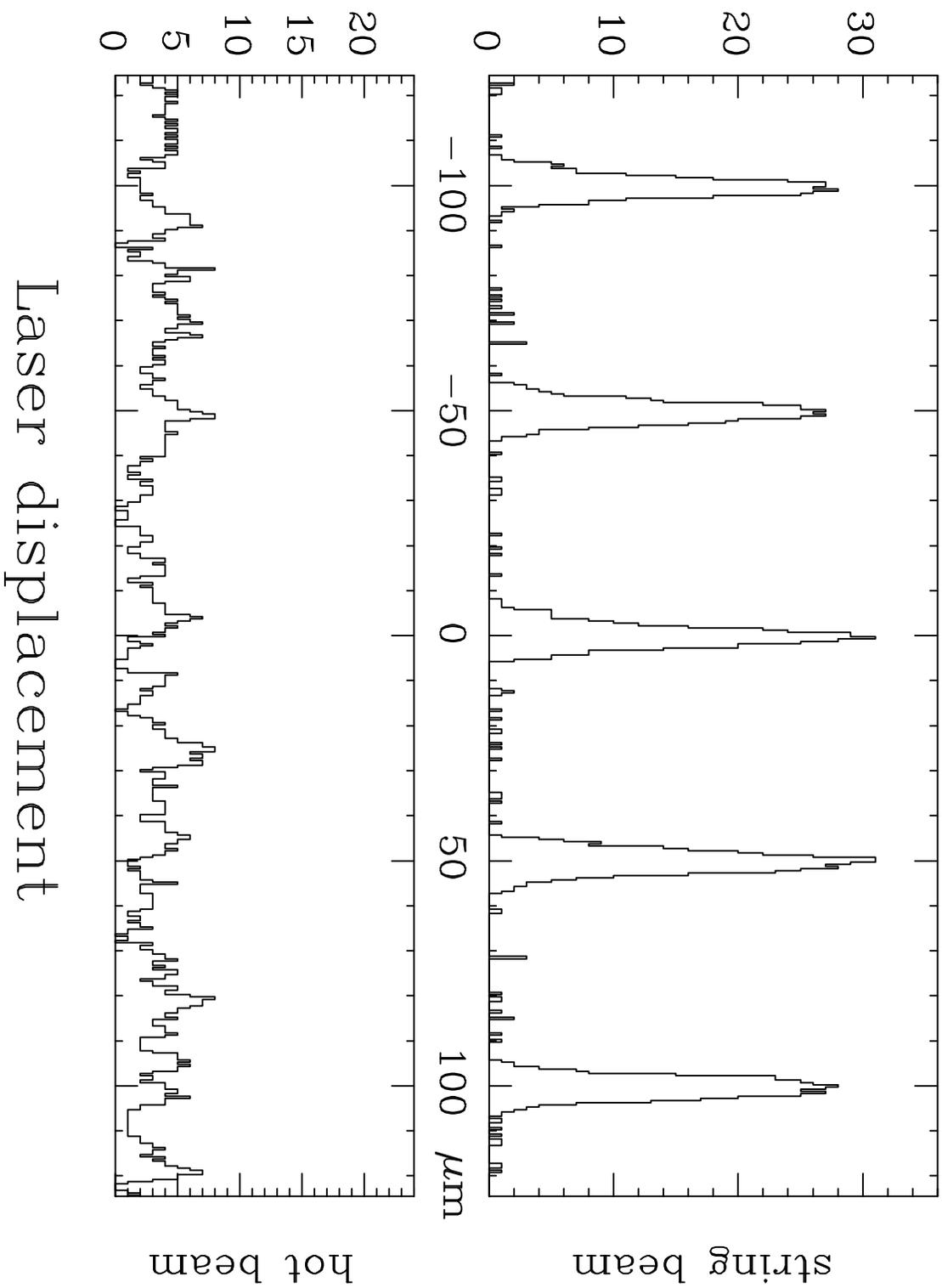
\end{document}